\documentstyle[12pt]{article}
\begin{document}
\hoffset = -1truecm
\voffset = -2truecm
\title{\bf
Light Propagation in Nonlinear Waveguide
and Classical Two-Dimensional Oscillator}
\author{A. Angelow\\
Institute of Solid State Physics, 72 Trackia Blvd., Sofia 1784, Bulgaria\\
E-mail: Angelow@bgcict.acad.bg\\}
\date{30 May 1997}

\maketitle

\begin{abstract}
The quantum optical problem of the propagation of electromagnetic
waves in a nonlinear waveguide is related to the solutions of the
classical nonstationary harmonic oscillator using the method of linear
integrals of motion [Malkin et.al., Phys Rev.\underline{2}D (1970) 1371].
An explicit solution of the classical oscillator with a varying frequency,
corresponding to the light propagation in an anisotropic waveguide is
obtained using the expressions for the quantum field fluctuations.\ \

Substitutions have been found which allow to establish connections of the
linear and quadratic invariants of Malkin et.al. to several types of
invariants of quadratic systems, considered in later papers. These
substitutions give the opportunity to relate the corresponding quantum
problem to that of the classical two-dimensional nonstationary oscillator,
which is physically more informative.

\end{abstract}
PACS number(s): 03.65Ca, 03.65Fd

\section{Introduction}\ \

The generation of subfluctuant (squeezed) and cofluctuant (covariance)
states of various systems has
stimulated a great deal of interest due to the important prospective
applications of these states in optical communications and super sensitive
detection systems. Many schemes of processes which generate the nonclassical
states have been suggested. A considerable attention has been paid to
the investigations of the nonclassical states of an oscillator with time-
dependent frequency or mass.
Recently, an oscillator with a linear sweep of restoring
force \cite{kn:agarwal}, and an oscillator which describes the quantum motion
in the Paul trap \cite{kn:brown},\cite{kn:paul} have been investigated.
F.L. Li and coauthors have also studied the squeezing property of an
oscillator with a time-dependent frequency \cite{kn:li1}-\cite{kn:li3}.
There are some other recent articles, considering explicit solutions of
quantum systems with quadratic Hamiltonians \cite{kn:brito}-\cite{kn:lai}.
We would like to note that two overcomplete families of solutions
for general quadratic Hamiltonians are built in \cite{kn:trifonov3,kn:holz}
by the method of linear invariants.\ \

One of the aims of this article is to establish a relationship between
the results in later works with those in
\cite{kn:trifonov1}-\cite{kn:holz}.\ \

Solving the Quantum -- Optical problem \cite{kn:angelow} for an anisotropic
waveguide, we have found a new solution of the equation for the classical
nonstationary harmonic oscillator with a time-dependent frequency:
\begin{equation}\label{eq:epsilon}
\ddot{\epsilon}+\Omega^2(t) \epsilon = 0.
\end{equation}
The second aim of this paper is to determine the classical nonstationary
harmonic oscillator, which corresponds to the quantum problem of the light
propagation in a second order nonlinear waveguide at the degenerated
parametric down conversion $\chi^{[2]}( 2\omega = \omega + \omega)$,
i.e. to find the time-dependent frequency
$\Omega(t)$ and the solution $\epsilon(t)$ of equation
(\ref{eq:epsilon}) corresponding to this case.\ \

In the next section we show that according to
the principle of correspondence between classical mechanics and
quantum mechanics there exist exactly $2s$ quantum linearly independent
invariants \cite{kn:trifonov3}. For quadratic Hamiltonians these $2s$
integrals of motion are linear in terms of $\hat q$ and $\hat p$,
therefore any problem with an one-dimensional quadratic
Hamiltonian (${\hat H}_{quad}$) is related to
the two-dimensional classical harmonic oscillator.
The found substitutions allow to transform some classical equations
considered in later papers, to the equation of the two-dimensional
nonstationary harmonic oscillator.
In the third section we apply the method of linear invariants to
anisotropic waveguides. In the fourth section we consider the
explicit solution for the classical nonstationary two-dimensional
harmonic oscillator for the anisotropic waveguide.

\section{Method of Linear Integrals of Motion}\ \

Let us consider a classical system with $s$ degree of freedom and let
$u=u(q_1,...q_s,$\, \,$p_1,...p_s,t)$ be one dynamic variable of this system.
Expressed in terms of Poisson brackets $\{,\}$, the full derivative of $u$
with respect to $t$ is:
\begin{equation}\label{eq:full_der}
{du\over dt}={{\partial u}\over {\partial t}} + \{ u,H_{class} \}=0 .
\end{equation}
By definition $u^{inv}$ will be an integral of motion (invariant) iff
\,\, ${du^{inv}/ dt}=0$, i.e.
\begin{equation}\label{eq:def_inv}
{{\partial u^{inv}}\over {\partial t}} + \{ u^{inv},H_{class} \}=0 .
\end{equation}
It is well known \cite{kn:nikolov} that there are $2s$ independent
integrals of motion, which are linear with respect to
$q_1,...q_s,p_1,...p_s$.
For example, one can choose the coordinates of the initial
point $(q_1(0),...q_s(0),p_1(0),...p_s(0))$
of the trajectory in the phase space:
\begin{eqnarray}\label{eq:ini_point}
{Q^{inv}_k} = q_k(0) &  \\ \nonumber
{P^{inv}_k} = p_k(0) & \qquad (k=1,...,s).
\end{eqnarray}
As far as there is a principle of correspondence between
classical and quantum mechanics, the analogy requires the existence of
$2s$ Hermitian operators - integrals of motion
for any quantum system, and the relevant equations of the quantum
invariants are \cite{kn:trifonov3}:
\begin{equation}\label{eq:ope_inv}
{{\partial {\hat I}^{inv}_{\nu}}\over {\partial t}} +
{1\over{i\hbar}} [{\hat I}^{inv}_{\nu},{\hat H}]=0,
\quad \nu=1,...,2s .
\end{equation}
The equations for the invariants (\ref{eq:ope_inv}) are different from the
Heisenberg equations:
${{d\hat A}\over {dt}}-{1\over {i\hbar}}[\hat A,\hat H]=0$.
The same difference exists in classical mechanics between (\ref{eq:def_inv})
and the Hamilton equations written in terms of Poisson brackets:
${{du}\over {dt}}-\{u,H_{class}\}=0$, $u=q_k,p_k$ \cite{kn:nikolov}.
The independent solutions of (\ref{eq:ope_inv}) for any quantum system are
also $2s$: ${\hat I}^{inv}_{\nu}(t)$ $=$ \, \, \,
${\hat U}(t) I^{inv}_{\nu}(0) {\hat U}^{\dagger}(t)$.
In particular, the invariants could be operators of the coordinates and
the moments at $t=0$,
\begin{equation}
{\hat Q}^{inv}_k(t)={\hat U}(t) {\hat q}_k(0) {\hat U}^{\dagger}(t)=
\sqrt{\hbar\over{2m\omega}}(\hat{A}^{\dagger}_k+\hat{A}_k) \nonumber
\end{equation}
\begin{equation}\label{eq:QP}
{\hat P}^{inv}_k(t)={\hat U}(t) {\hat p}_k(0) {\hat U}^{\dagger}(t)=
i\sqrt{\hbar m\omega\over{2}}(\hat{A}^{\dagger}_k-\hat{A}_k) ,
\end{equation}
where for the further convenience they are presented as real
and imaginary parts of $s$ non-hermitian operators $\hat{A}_k$
( m and $\omega$ are constants with dimension of mass and frequency
respectively ).\ \

If $\hat A$ is an invariant, it will be easy to verify that any function
presented as power series $F(\hat A)$ is an invariant too.
The invariants  for the general quadratic quantum system with
$\hat H_{quad}$ are
published in \cite{kn:trifonov3,kn:holz}, and it is worth noting
that they reproduce any quadratic or higher degree invariants.
Generally, using the formula
\begin{equation}\label{eq:invariant2}
I_g(t)=\alpha_{1} \hat{A}^{\dagger 2} + \alpha_{2}\hat{A}^{\dagger}\hat{A}+
\alpha_{3} \hat{A}^2 + \alpha_{4} \hat{A}^{\dagger} +
\alpha_{5} \hat{A} + \alpha_{6} ,
\end{equation}
any quadratic integrals of motion for $\hat H_{quad}$  could be
built up with the help of $\hat{A}(t)$.
Here, $\alpha_i$, $i=1,..6$ are time-independent coefficients.
See \cite{kn:lo2,kn:li1,kn:ji,kn:lai,kn:gunther} where the linear
invariants from \cite{kn:trifonov2,kn:trifonov3} are not mentioned
at all.\ \

Following \cite{kn:trifonov1,kn:trifonov2}, we will shortly recall
the method of linear invariants and will apply it in the next
sections in the case of an one-mode waveguides to find the explicit
solution of the two-dimensional harmonic oscillator.
%
%
Without losing generality, let us consider the one-dimensional
case (i.e. $s=1$) of a nonstationary quantum system described
by a quadratic Hamiltonian (the general case is considered in
the first paper \cite{kn:trifonov3}, eq.(1)\,)
\begin{equation}\label{eq:HH}
{\hat H}= a(t){\hat p}^2 + b(t)[p,q]_+ + c(t){\hat q}^2 +d(t)p+ e(t)q+ f(t),
\end{equation}
where $a(t),b(t),c(t),d(t),e(t),f(t)$ are arbitrary real functions of $t$,
and $[\quad,\quad]_+$ denotes an anti-commutator.
For this Hamiltonian (see \cite{kn:trifonov1}, eq.(3))
we can construct $s=1$ non-Hermitian linear integral (\ref{eq:QP})
of motion (and its Hermitian conjugate, respectively) $$
{\hat A} ={i\over\sqrt{\hbar a(t)}}\left\{ {a(t)\epsilon(t) \hat p} +
\left [ b(t)\epsilon(t) -
{\dot{\epsilon}(t)\over 2} -{1\over 4}{{\dot{a}(t)}\over a(t)}
\epsilon(t)\right ] {\hat q}\right\} +\delta(t),
$$
\begin{equation}\label{eq:invariants}
{\hat A}^\dagger=-{i\over\sqrt{\hbar a(t)}}\left\{
{a(t)\epsilon^{*}(t) \hat p} + \left [ b(t)\epsilon^{*}(t) -
{\dot{\epsilon^{*}}(t)\over 2} -{1\over 4}{{\dot{a}(t)}\over a(t)}
\epsilon^{*}(t)\right ] {\hat q}\right\} +\delta(t) .
\end{equation}
Here $\epsilon(t)$ is a complex function satisfying
(\ref{eq:epsilon}) and $\delta(t)$ is expressed in terms of
$\epsilon$ and $\dot{\epsilon}$ similar to \cite{kn:trifonov3}. Since
we want to deal with hermitian  Hamiltonians (i.e. real $a,b,c,d,e,f$) let
us introduce real functions $\epsilon_1 (t), \epsilon_2(t)$ and
$\delta_1(t),\delta_2(t)$ so that
\begin{equation}\label{eq:eps-del}
\epsilon(t)=\epsilon_1(t)+i\epsilon_2(t) ,\quad
\delta(t)=\delta_1(t)+i\delta_2(t).
\end{equation}\ \
We shall show that the classical nonstationary two-dimensional harmonic
oscillator which corresponds to this general quadratic system with one
degree of freedom and quadratic Hamiltonian (\ref{eq:HH})
has a Lagrangian
\begin{equation}\label{eq:lagrangian}
{L={{m\over2}(\dot{\epsilon}_1^2+\dot{\epsilon}_2^2) -
{m\over2}{\Omega^2(t)}(\epsilon_1^2+\epsilon_2^2)}} .
\end{equation}
The time-dependent frequency
$\Omega =\Omega(t)$ of this classical harmonic oscillator
achieves different forms for the different quantum systems.
We shall recall the general form \cite{kn:trifonov1} of this frequency
and we will apply it to find the explicit expression for the special
case of the quantum propagation of light in a nonlinear waveguide.\ \

Let us construct the second type Lagrange equations \cite{kn:nikolov}
for this two-dimensional nonstationary harmonic oscillator ($m=1$;
the case with a time-dependent mass $M(t)$ could be reduced to this one by
a transformation of time, see for example eq.(122) of the second paper in
\cite{kn:trifonov3}):
\begin{equation}\label{eq:lag-eq}
{d\over dt} {\partial L\over \partial {\dot{\epsilon}}_{\nu}} -
{\partial L\over \partial \epsilon_{\nu}}=0 , k=1,2 .
\end{equation}
%

These two real equations can be written as
\begin{equation}\label{eq:xx}
{\ddot \epsilon_{\nu}} + \Omega^2 (t)\epsilon_{\nu} =0 , \quad \nu=1,2
\end{equation}
which differ from the equation for the simple harmonic oscillator by the
fact that the frequency is time-dependent $\Omega=\Omega(t)$.\ \

Without losing generality in deriving the expression for the frequency
$\Omega(t)$,
we assume that $d(t)=e(t)=f(t) =0= \delta_1(t)= \delta_2(t)$.
Since the operators (\ref{eq:invariants}) are integrals of motion,
they obey the (necessary and sufficient) conditions (\ref{eq:ope_inv}).
Hence
\begin{eqnarray}\label{eq:eq-motion}
& {\partial\over \partial t}\left (
{i\over\sqrt{\hbar a(t)}}\left\{ {a(t)\epsilon(t) \hat p} +
\left [ b(t)\epsilon(t) -
{\dot{\epsilon}(t)\over 2} -{1\over 4}{{\dot{a}(t)}\over a(t)}
\epsilon(t)\right ] {\hat q}\right\}
\right )-                                                     & \\ \nonumber
& {i\over \hbar}\left[
{i\over\sqrt{\hbar a(t)}}\left\{{a(t)\epsilon(t)\hat p} +
\left[ b(t)\epsilon(t)-
{\dot{\epsilon}(t)\over 2}-{1\over4}{{\dot{a}(t)}\over a(t)}
\epsilon(t)\right]{\hat q}\right\},
a(t){\hat p}^2+b(t)[p,q]_+ +c(t){\hat q}^2
\right]=0 \, .                                                & \nonumber
\end{eqnarray}
It reduces to
\begin{eqnarray}\label{eq:eq-motion1}
&
\left\{ 0\cdot \hat p + \left[
-\ddot{\epsilon} + 0\cdot\dot{\epsilon} +
\left( -{ 4 a(t)c(t) - 2{\dot{a}(t)\over a(t)}b(t) -
{\ddot{a}(t)\over 2 a(t)} + {3 {\dot{a}(t)}^{2}\over 4 {a(t)}^{2}}
+ 4b(t)^{2} + 2\dot{b}(t)}\right)\epsilon
\right] \cdot \hat q\right\}   \nonumber \\
&          \qquad \qquad \qquad =0 \, .
\end{eqnarray}
From this equation we see, that $\hat A$ and ${\hat A}^\dagger$
will be the quantum invariants for the system with the general
nonstationary Hamiltonian (\ref{eq:HH}), if the frequency $\Omega(t)$
is connected with time-dependent coefficients of the Hamiltonian
in the following way \cite{kn:trifonov1}:
\begin{equation}\label{eq:omega1}
{\Omega}^{2}(t)= {4 a(t)c(t) + 2{\dot{a}(t)\over a(t)}b(t) +
{\ddot{a}(t)\over 2 a(t)} - {3 {\dot{a}(t)}^{2}\over 4 {a(t)}^{2}}
- 4b(t)^{2} - 2\dot{b}(t)},
\end{equation}
and $\epsilon_1$ and $\epsilon_2$ are solutions of classical
Lagrange equations (\ref{eq:lag-eq}) for the two-dimensional
nonstationary oscillator, or respectively, the solution
$\epsilon=\epsilon_1 + i\epsilon_2$  of the classical
complex nonstationary harmonic oscillator.
This solution can be represented in the form
\begin{equation}\label{eq:epsilon1}
\epsilon(t) = \rho(t) \ e^{i\int_{0}^{t}{d\tau\over \rho^2(\tau)}},
\end{equation}
where $\rho (t) \equiv |\epsilon(t)|$ \cite{kn:trifonov3}.\ \

The efforts of many investigations have been directed to the quadratic
quantum system whose evolutions are described by a classical differential
equation like the equation (\ref{eq:epsilon}).
In the literature several equations
\cite{kn:lewis,kn:lo1,kn:lo2,kn:ji,kn:piza}
connected with the harmonic oscillator or with its generalizations
are published.
In Table 1. we give the substitutions which transform these equations
to the classical nonstationary complex oscillator
(or the same -- to the two-dimensional oscillator).
The first column presents the references, where the equations are
taken from.
More details on these equations and the relevant substitutions are
presented in Appendix 1.\ \

We would like to note another independent approach to this
method presented by Toledo de Piza \cite{kn:piza}, where the
evolution of the quantum system in pure and mixed states is described by
classical Hamilton equations. In the case of pure states, such two
systems of Hamilton equations for two pairs of real physical
parameters \{$q,p$\} and \{$\sigma,\Pi$\} have been derived
earlier in \cite{kn:cofluctuant} to describe completely the class
of the Schr{\" o}dinger minimum uncertainty states. The fourth
substitution in Table 1 shows the explicit connection between the
classical Hamilton equation and the two-dimensional harmonic oscillator,
see Appendix 1.\ \

The method of linear integrals of motion is very powerful
and can be used in different directions:
defining the evolution of each quantum state, transition probabilities,
Berry phase etc.
\cite{kn:trifonov2,kn:trifonov3,kn:brown,kn:seleznyova,kn:cofluctuant}.\ \

With these comments on the general quadratic Hamiltonian and the
two-dimensional harmonic oscillator we have shortly made a resume of
the method of linear invariants, developed in
series of papers \cite{kn:trifonov1,kn:trifonov2,kn:trifonov3,kn:holz};
for any quantum system, described by Hamiltonian in form (\ref{eq:HH}),
there exists a classical two-dimensional isotropic nonstationary harmonic
oscillator (or the same -- complex nonstationary harmonic oscillator
(\ref{eq:epsilon})) with a Lagrangian (\ref{eq:lagrangian}) and equations
of motion (\ref{eq:lag-eq}) with nonstationary frequency $\Omega(t)$
(\ref{eq:omega1}).\ \

\section{Light propagation in anisotropic waveguide and
quadratic Hamiltonians}\ \

The problem of a propagation of light from the quantum-mechanical
point of view has been investigated in series of papers
\cite{kn:walls1,kn:drummond,kn:angelow}.

In \cite{kn:angelow} it has been shown that the Hamiltonian,
\begin{equation}\label{eq:hamiltonian1}
\hat{H}={\hbar \omega }(\hat{a}^{\dagger}\hat{a}+{1\over 2})\ +\ {\hbar s}\
[e^{i2\omega t}\hat{a}\hat{a}+e^{-i2\omega t}\hat{a}^\dagger\hat{a}^\dagger]
\end{equation}
describes the light propagation in the nonlinear $Ti:LiNbO_3$ waveguide
in the case of a degenerated  parametric down conversion.
In the above mentioned paper it has been  found an explicit form
of Heisenberg equations, and are also calculated quantum fluctuations
of the electromagnetic field $\sigma_{q}(t)$, $\sigma_{p}(t)$
and their cofluctuation $c_{qp}(t)$ (the three independent second moments
in probability theory \cite{kn:cofluctuant} ):
\begin{eqnarray}\label{eq:cov}
\sigma_{q}^{2}(t) &=& {\hbar\over2\omega}\left(1+2\sinh^{2}(st)-
\sinh(2st)\sin(2\omega t)\right) \label{eq:sq2}, \nonumber \\
\sigma_{p}^{2}(t) &=& {\hbar\omega\over2}\left(1+2\sinh^{2}(st)+
\sinh(2st)\sin(2\omega t)\right) \label{eq:sp2}, \\
{c_{qp}^2(t)} &=& {{\hbar^2\over4}\ \sinh^2(2st)\
\cos^2(2\omega t)} . \nonumber
\end{eqnarray}\ \
The Hamiltonian (\ref{eq:hamiltonian1}) has the same form in volume
nonlinear materials \cite{kn:yariv} though with different
coefficients since in our case the squeezed parameter $s$
depends on the waveguide parameters and phase-matching
conditions for this experimental geometry \cite{kn:angelow}.\ \

Now we are going to apply this method in the opposite direction; since
we have solved the quantum problem for the nonlinear anisotropic waveguide
with particular Hamiltonian of type (\ref{eq:HH}) we will now find
the explicit form of the frequency $\Omega(t)$ and solutions
$\epsilon_1(t)$ and $\epsilon_2(t)$ for this harmonic
oscillator.\ \

In order to define the time-dependent coefficients
$a(t),b(t)$ and $c(t)$ we express the Hamiltonian (\ref{eq:hamiltonian1})
in terms of canonical operators $\hat{q}$ and $\hat{p}$
in the Schr\"odinger  picture,
\begin{equation}\label{eq:hamiltonian3}
\hat{H}=(1-{2s\over \omega}\cos(2\omega t))\hat{p}^{2}
+(-{2s \omega} \sin(2\omega t))[\hat{p},\hat{q}]_{+}+
({\omega^{2}}-{2s\over \omega}\cos(2\omega t))\hat{q}^{2} .
\end{equation}
Hence, for the time-dependent coefficients of (\ref{eq:HH})
we obtain,
$$
a(t)= {1\over 2}-{s\over \omega}\cos(2\omega t),
$$
\begin{equation}\label{eq:b}
b(t)= - s \sin(2\omega t),
\end{equation}
$$
c(t)= {\omega^{2}\over2}+{s \omega}\cos(2\omega t),
$$\ \

Using these results for the quantum system
describing the light propagation in an anisotropic waveguide,
we are able to express in explicit form the
solution of equation (\ref{eq:epsilon}) and the frequency $\Omega(t)$.

Many authors in their papers, concerning nonstationary harmonic oscillator
for different cases of time-dependent frequencies
\cite{kn:lo1}-\cite{kn:ji},
have described frequencies which are particular cases of general formulae
received in \cite{kn:trifonov2,kn:trifonov3}.
Besides, up to now no explicit solution of equation
(\ref{eq:epsilon}) has been found for the case of parametric
down-conversion in a nonlinear waveguide, which is described by the
Hamiltonian (\ref{eq:hamiltonian1}),(\ref{eq:hamiltonian3}).
The subject of the following
section is to consider that frequency for the classical equation of
the harmonic oscillator, whose solutions determine the quantum dynamics of
the electromagnetic field for such waveguides.

\section{Explicit solutions}\ \

In this section we shall derive the formula for the nonstationary frequency,
with the help of which we shall give the explicit solution of the equation
(\ref{eq:epsilon}) (respectively (\ref{eq:xx})\,) in the case of an
anisotropic waveguide.
Using equations (\ref{eq:b}) and substituting in the equation
(\ref{eq:omega1}) we obtain the frequency $\Omega(t)$ in the following
explicit form:
\begin{equation}\label{eq:omega2}
{\Omega}^{2}(t)=  {\omega}^{2}-4s^{2} +
4 s \omega \cos(2\omega t)+
{4s{\omega}^{2}  \cos(2\omega t)-
8 s^{2} {\omega}\sin^{2}(2\omega t)
\over \omega - 2s \cos(2\omega t)} -
{12s^{2}{\omega}^{2} \sin^{2}(2\omega t)
\over (\omega - 2s \cos(2\omega t))^{2}} .   \nonumber
\end{equation}

It is very important in our consideration that we have solved the
quantum problem in an independent way, i.e. we have determined the
evolution of the operators $\hat p$ and $\hat q$ and their second
moments (\ref{eq:cov}) by a direct calculation \cite{kn:angelow} .
On the other hand, a connection exists between the modulus
$\rho(t)=|\epsilon(t)|$ and the quantum mechanical fluctuations
of $\hat q$ and $\hat p$, as shown in \cite{kn:trifonov2}.
In order to calculate the fluctuations we express $\hat q$ and
$\hat p$ in terms of
linear invariants $\hat{A}^{\dagger}(t)$ and $\hat{A}(t)$
(\ref{eq:invariants}) and take their mean value with respect to
the eigenstates of the linear invariants
($\hat{A}(t) |z;\alpha > = z |z;\alpha >$ ). The result is:
\begin{eqnarray}\label{eq:cov1}
\sigma_{q}^2(t) &=& \hbar \,\,a(t) \,\rho^2(t), \nonumber \\
\sigma_p^2(t)   &=& {\hbar\over a(t)}\left[{1\over{4\rho^2(t)}}+
\left( b(t)\rho(t)-{\dot{\rho}(t)\over2}-
{\dot{a}(t)\over {4a(t)}}\rho(t)\right)^2\right], \\
c_{qp}^2(t)     &=& \hbar^2 \rho^2(t) \left( b(t)\rho(t)-
{\dot{\rho}(t)\over2}-{\dot{a}(t)\over {4a(t)}}\rho(t)\right)^2. \nonumber
\end{eqnarray}\ \
The last formula in (\ref{eq:cov1}) presents the third independent second
moment (cofluctuation) $c_{qp}(t)$ expressed in terms of the modulus
$\rho(t)\equiv|\epsilon(t)|$.
This formula for $c_{qp}(t)$ has been derived on the base of the
Schr{\" o}dinger uncertainty relation \cite{kn:schrodinger}
(see Appendix 2 ).\ \

Now we are able to define $\rho(t)$ as a function of the time and
parameters $\omega$ and $s$, using the expression for $\sigma_q(t)$ from
(\ref{eq:cov}) and (\ref{eq:cov1}) to eliminate it.
Once we have an expression for  $\rho(t)$ we can get the final
solution $\epsilon(t)$ by the general formula (\ref{eq:epsilon1}).
Thus, we have obtained the explicit solution of the classical equation
for the nonstationary harmonic oscillator (\ref{eq:epsilon}) with
frequency (\ref{eq:omega2}):
\begin{eqnarray}\label{eq:epsilon2}
&\epsilon(t)&=
\sqrt{\left(1+2\sinh^{2}(st)-\sinh(2st)\sin(2\omega t)\right)
\over \omega - 2s \cos(2\omega t)} \times \\
&           &
\times exp\left[i\int_{0}^{t}{{\omega - 2s \cos(2\omega \tau)
\over 1+2\sinh^{2}(s\tau)-\sinh(2s\tau)\sin(2\omega \tau)}
d\tau}\right] . \nonumber
\end{eqnarray}
The eigenstates of the linear invariants (\ref{eq:invariants}) satisfy
${\hat A} |z;\alpha> = z |z;\alpha>$, at the same time they are
Schr{\" o}dinger Minimum Uncertainty States
(i.e. $|SMUS>=|z;\alpha>$) as shown
in \cite{kn:cofluctuant}. Because $|SMUS>$ form an over-complete
system of eigenfunctions
\begin{equation}\label{eq:completeness}
{1\over \pi}\int{|z;\alpha ><\alpha ;z|}d^2 z = 1 ,\qquad
d^2 z \,\equiv\, d(Re z) d(Im z),
\end{equation}
and remain stable in the time evolution, governed by Hamiltonians
(\ref{eq:HH}) \cite{kn:cofluctuant}, the class of two-dimensional
isotropic oscillators (i.e. $\epsilon_{1,2}(t)$) determines the dynamics
of any quantum state $\psi(t)$; in particular it determines the propagation
of light in nonlinear waveguides with second order nonlinearity
$\chi^{(2)}( \omega_1 = \omega_2 + \omega_3 )$ (see \cite{kn:angelow}).

\section{Conclusion}\ \

It is well known that solutions for the equation of time-dependent
harmonic oscillator are found only in few particular cases \cite{kn:kamke}.
The present work  gives one more contribution to the class of solutions
of the particular oscillator's equation.\ \

We have found a solution of the equation for the classical
nonstationary harmonic oscillator with a time-dependent frequency
(\ref{eq:omega2}) in integral form (\ref{eq:epsilon2}). We have shown
that the quantum-optical problem for the propagation of electromagnetic waves
in the nonlinear waveguide \cite{kn:angelow} is related to the solutions of
the classical nonstationary harmonic oscillator. This method could be used
for obtaining other solutions of the classical harmonic oscillator
with different time-dependent frequency, if we have already determined
the fluctuations $\sigma_{q}$ and $\sigma_{p}$ by some independent
methods.\ \

We have also found the relations between the general method of
linear invariants and the equations given in the second column
in Table 1. This method gives us the substitutions which transform
the classical equations to the physically more informative equation of
the two-dimensional nonstationary harmonic oscillator.
We should also note that for any quantum system described by a
nonstationary quadratic Hamiltonian, there is not only a correspondence
to a complex oscillator, but also a correspondence to a real two-dimensional
isotropic oscillator with nonstationary frequency.\ \

Using the Schr{\" o}dinger uncertainty relation (discovered in
1930 year) we have obtained the evolution of the cofluctuation $c_{qp}$,
expressed by the solutions of the equation for the classical nonstationary
harmonic oscillator. The cofluctuation plays significant role in the so
called Cofluctuant States, which differ from widely studied Coherent and
Squeezed States.\ \

\section{Acknowledgments}\ \

This work was partly supported by Bulgarian Scientific Foundation,\\
grant number F-81 and F-559.\ \

The author thanks D.A. Trifonov for the helpful comments.\ \

\section*{Appendix 1}\ \

Here we shall prove the substitution from Table 1, which
gives us the connection between the equation of the two-dimensional
nonstationary harmonic oscillator and other classical equations,
describing the quantum evolution of a system with
quadratic Hamiltonians.\ \

\ \

First, we shall show that the Ricatti equation for $c_1(t)$ from
\cite{kn:lo2} can be transformed to equation (\ref{eq:lag-eq}),
and vice versa, with the help of the second substitution from Table 1.
Here we shall use the same functions of time $c_2(t)$, $c_3(t)$,
${a_1}^{\prime}(t)$, ${a_2}^{\prime}(t)$ and ${a_3}^{\prime}(t)$
as in \cite{kn:lo2}:
\begin{equation}\label{eq:c2c3}
c_2(t)= \int_0^t{[{a_2}^{\prime}(\tau)+
{a_3}^{\prime}(\tau)c_1(\tau)]}d\tau \,\,\,\,\,\,\,
c_3(t)= \int_0^t{{a_3}^{\prime}(\tau) e^{c_2(\tau)}}d\tau .
\end{equation}
The time-dependent coefficients in the Hamiltonian from \cite{kn:lo2}
${a_1}^{\prime}(t)$, ${a_2}^{\prime}(t)$ and ${a_3}^{\prime}(t)$ are real,
and are connected with our coefficients as follows:
\begin{eqnarray}\label{eq:a1a2a3}
\nonumber
{a_1}^{\prime}(t)=-2 c(t)  & \\
{a_2}^{\prime}(t)=-4 b(t)  & \\ \nonumber
{a_3}^{\prime}(t)=-2 a(t). & \nonumber
\end{eqnarray}
With these notations we can express the common solution of the
complex equation for a nonstationary harmonic oscillator in the
following way
\begin{equation}\label{eq:substitution}
\epsilon(t)= - {e^{-{c_{2}(t)\over 2}}\over \sqrt{m_0 \omega_0 a_3(t)}}
(m_0 \omega_0 c_3(t) + i ) ,
\end{equation}
where to simplify expressions we have used $c_2(t)$ and $c_3(t)$ from
\cite{kn:lo2}.
After some calculations the second derivative of
(\ref{eq:substitution})  becomes:
\begin{eqnarray}\label{eq:ddot_e1}
& \ddot{\epsilon}={e^{-{c_{2}(t)\over 2}}\over \sqrt{m_0 \omega_0
{a_3}^{\prime}(t)}}{(m_0 \omega_0 c_3(t) + i)}\times \\
& {\left[ {a_3}^{\prime} (-{a_2}^{\prime} c_1 -{a_3}^{\prime} {c_1}^2 +
\dot{c_1} ) -{a_3}^{\prime} {a_1}^{\prime}
+ {a_3}^{\prime} {a_1}^{\prime} -{1\over2}{\dot{{a_3}^{\prime}}\over
{a_3}^{\prime}} {a_2}^{\prime}
-{3\over4}{{{\dot{{a_3}^{\prime}}}^2}\over {{a_3}^{\prime}}^2}
+{1\over2}{\ddot{{a_3}^{\prime}}\over {a_3}^{\prime}}
-{{{a_2}^{\prime}}^2 \over 2}
+{\dot{{a_2}^{\prime}}\over 2} \right]} . \nonumber
\end{eqnarray}
We have obtained exactly the frequency
$\Omega^2(t)$ (\ref{eq:omega1}) expressed through
${a_1}^{\prime}(t)$, ${a_2}^{\prime}(t)$ and ${a_3}^{\prime}(t)$
from \cite{kn:lo2}.
Taking into account the relations (\ref{eq:substitution}) and
(\ref{eq:a1a2a3}) we receive:
\begin{equation}\label{eq:ddot_e3}
\ddot{\epsilon}= {- \epsilon}
{\left[ {a_3}^{\prime} (-{a_1}^{\prime} -{a_2}^{\prime} c_1 -
{a_3}^{\prime} {c_1}^2 + \dot{c_1} ) +\Omega^2(t) \right]} . \nonumber
\end{equation}
As far as the Ricatti equation (eq.(9) from \cite{kn:lo2}) should
be fulfilled
\begin{equation}\label{eq:ricatti}
\dot{c_1} = {a_1}^{\prime} + {a_2}^{\prime} c_1 + {a_3}^{\prime} {c_1}^2
\end{equation}
the expression in parentheses is zero. As a result, we obtain the
equation (\ref{eq:epsilon}). Presenting $\epsilon(t)$ as a
real and an imaginary part, we obtain the  classical system of equations
(\ref{eq:lag-eq}) of a two-dimensional nonstationary harmonic oscillator.
These two equations could also be derived directly, working separately
with the real and imaginary parts of the second substitution from Table 1.
The opposite is also true; if we want eq. (\ref{eq:epsilon}) to be
fulfilled then using (\ref{eq:ddot_e3}) we shall receive
the Ricatti equation for $c_1(t)$.\ \

\ \

Second, we shall show that the equation (8)  for $f(t)$ from \cite{kn:ji}
\begin{equation}\label{eq:epsilon4}
\ddot{f}+{\dot{m(t)}\over m(t)}\dot{f}+\omega^2(t) f = 0
\end{equation}
can be also transformed to equation (\ref{eq:epsilon}) or (\ref{eq:lag-eq}).
Using the well known substitution \cite{kn:kamke}
\begin{equation}\label{eq:ft}
f(t)={\epsilon(t)\over \sqrt{m(t)}} ,
\end{equation}
we again obtain the equation
\begin{equation}\label{eq:ji_eq}
\ddot{\epsilon} +
{\left[ \omega^2(t)+{1\over4}{\dot{m(t)}^2\over {m(t)^2}}-
{1\over2}{\ddot{m(t)}\over m(t)} \right] } {\epsilon} = 0
\end{equation}
with a frequency $\Omega(t)$, which is a particular case of
(\ref{eq:omega1}); the coefficient in front of the anti-commutator
$[\hat{p},\hat{q}]_+$ in first article in \cite{kn:ji} is zero,
which simplifies the expression for the frequency.
Note that $\epsilon(t)$ has no imaginary part, i.e. equations
(\ref{eq:lag-eq}) are identical and the oscillator becomes
one-dimensional.\ \

\ \

Third, we shall show that Hamilton equations (32) from \cite{kn:cofluctuant}
for the pair of canonical variables $\sigma(t)\equiv \sigma_q$
and $\Pi\equiv c / \sigma_p$, where $c\equiv \sigma_{qp}$ is the
third independent second momentum -- the so called  cofluctuation. Namely,
\begin{eqnarray}\label{eq:ham_eq}
& \dot{\sigma}= & {\partial <H_1>\over \partial \sigma}=
2 b \sigma + 2 a \Pi \\ \nonumber
& \dot{\Pi}=    & -{\partial <H_1>\over \partial \Pi}=
- 2 c \sigma - 2 b \Pi + {\hbar^2 a\over 2}{1\over {\sigma}^3}
\end{eqnarray}
can also be transformed to the equation (\ref{eq:epsilon}) with
the help of the following classical Hamiltonian:
\begin{equation}\label{eq:H_1}
H_1(\sigma(t),\Pi(t))=a(t)\left( {1\over{4 \sigma^2}}+ \Pi^2\right) +
2 b(t)\sigma\Pi + c(t) \sigma^2  .
\end{equation}
This classical Hamilton function is the second part of the quantum mean
value of the Hamiltonian (\ref{eq:HH}) taken in Schr{\" o}dinger
minimum uncertainty states $|z;\alpha>$:
\begin{equation}\label{eq:mean_HH}
<\hat{H}(\hat{p},\hat{q})>=\mathaccent126H_0(p,q) +
\mathaccent126H_1(\sigma,\Pi) ,
\end{equation}
where $p=<\hat{p}>,q=<\hat{q}>$.
If we differentiate the substitution for $\sigma$ from Table 1, and replace
it in the first Hamilton equation (\ref{eq:ham_eq}), ( here
$\Pi=\Pi(\epsilon,\dot{\epsilon})$ is implicit function of $t$ )
we receive a first order differential equation for $\epsilon(t)$.
Differentiating it again and replacing $\dot{\Pi}$ from the second
Hamilton equation (\ref{eq:ham_eq}), we obtain
\begin{equation}
\ddot{\epsilon}+ ({4 a(t)c(t) + 2{\dot{a}(t)\over a(t)}b(t) +
{\ddot{a}(t)\over 2 a(t)} - {3 {\dot{a}(t)}^{2}\over 4 {a(t)}^{2}}
- 4b(t)^{2} - 2\dot{b}(t)})\epsilon+
\end{equation}
\begin{equation}
i{
{\dot{\epsilon}^*\epsilon-\dot{\epsilon}\epsilon^*}
\over{{\epsilon^*}^2\epsilon} } - {1\over {{\epsilon^*}^2\epsilon}}-
{1\over
{\epsilon^3 e^{-4i\int_{0}^{t}{d\tau \over {\epsilon^*\epsilon}} }} }
 =0 ,
\end{equation}
where the substitution for $\Pi$ from Table 1 has been also used. Taking
into account that the Wronskian for eq. (\ref{eq:epsilon}) is
$W(\epsilon,\dot{\epsilon})=
{\dot{\epsilon}\epsilon^* -\epsilon\dot{\epsilon}^*} =2i$, and
relation ${\epsilon^*}^2=\epsilon^2
e^{-4i\int_{0}^{t}{d\tau\over \epsilon^*\epsilon}}$, we obtain
(\ref{eq:epsilon}). Therefore, we have received the same frequency
$\Omega(t)$ as in the equation (\ref{eq:omega1}).\ \

It is easily to show that classical equations
$\dot{q}= {\partial \mathaccent126H_0(p,q)\over \partial p}$ and
$\dot{p}= -{\partial \mathaccent126H_0(p,q)\over \partial q}$
could also be transformed to (\ref{eq:epsilon}) or to Lagrange equations
(\ref{eq:lag-eq}) with the following substitutions
\begin{equation}\label{eq:class_qp}
q=\sqrt{\hbar a} \,\,2 Re(z\epsilon^*) \quad \quad \quad \quad
\end{equation}
$$
p=-\sqrt{\hbar \over a} \,\,2 Re\left[ z(b\epsilon-{\dot{\epsilon}\over 2}-
{1\over 4}{\dot{a}\over a}\epsilon)^*\right] ,
$$
where the complex $z$ are the eigenvalues of Schr{\" o}dinger
minimum uncertainty states (see (\ref{eq:completeness})\,).\ \

\section*{Appendix 2}\ \

Here we calculate the cofluctuation (third eq. in (\ref{eq:cov1}) ).
By definition, the third independent second moment $c_{qp}(t)$ in
the probability theory is
\begin{equation}\label{eq:cov2}
c_{qp}(t)={1\over2}<\hat{q}\hat{p}+\hat{p}\hat{q}>-
<\hat{q}><\hat{p}>.
\end{equation}
We can directly calculate it as we have done for $\sigma_q$ and
$\sigma_p$ in (\ref{eq:cov1}). This
derivation involves again the use of the Wronskian
$W(\epsilon,\dot{\epsilon})$ and formula
$
\epsilon(t) \dot{\epsilon}^*(t) = \rho(t) \dot{\rho}(t) - i ,
$
following from (\ref{eq:epsilon1}).
But there is an easy way (using a physical argument - uncertainty
principle) to derive the cofluctuation, since we have already expressed
the quantum fluctuations by $\rho(t)$. As we have
mentioned in (\ref{eq:completeness}), the eigenstates of
the linear invariants (\ref{eq:invariants}) are Schr{\" o}dinger
minimum uncertainty states $|z;\alpha >$.
The Schr{\" o}dinger uncertainty relation \cite{kn:schrodinger}
\begin{equation}\label{eq:schrodinger}
\sigma_{q}^{2} \ \sigma_{p}^{2}\geq
{1\over4}|[\hat{q},\hat{p}]|^{2} +c_{qp}^{2},
\qquad [\hat{q},\hat{p}]=i\hbar
\end{equation}
becomes equality for these states $|z;\alpha >$. Expressing $c_{qp}^{2}$
from this relation, we obtain
\begin{equation}\label{eq:schrodinger1}
c_{qp}^{2} = \sigma_{q}^{2} \ \sigma_{p}^{2} - {\hbar^2\over4}.
\end{equation}
Taking the fluctuations $\sigma_{q}$ and $\sigma_{p}$ from (\ref{eq:cov1}),
the final result for the cofluctuation becomes:
\begin{equation}\label{eq:11/11/96p15}
c_{qp}^{2} = \hbar^2 \rho^2(t) \left( b(t)\rho(t)-
{\dot{\rho}(t)\over2}-{\dot{a}(t)\over {4a(t)}}\rho(t)\right)^2.
\end{equation}

\newpage
\begin{center}
Table 1.
\end{center}
\vspace{5mm}

\begin{center}
\begin{tabular}{|l|l|l|l|} \hline
  & Refer.     &   Equations                    & Substitution to get eq.
\, $\ddot{\epsilon}+\Omega^2\epsilon = 0;$ \\
  &  $\,\,\,$  &                                & $\qquad \qquad
\quad \qquad \qquad \qquad \epsilon=\epsilon_1+i\epsilon_2$ \\ \hline
  &  $\,\,\,$  &                                &             \\
1 & \cite{kn:lewis,kn:trifonov3}
               & $\ddot{\rho}-{1\over \rho^3}+\Omega^2(t) \rho =0$
                                                & $\rho(t)=\epsilon(t)
\, e^{i\int_{0}^{t}{d\tau\over \epsilon^*(\tau)\epsilon(\tau)}}$
                        \quad follows from (\ref{eq:epsilon1}) \\
  &  $\,\,\,$  &                                &             \\ \hline
  &            &                                &             \\
2 & \cite{kn:lo2}
               & $\dot{c_1} = {a_1}^{\prime} + {a_2}^{\prime} c_1 +
{a_3}^{\prime} {c_1}^2                        $ & $\epsilon(t)=-
{e^{-{1\over2}{{\int_0^t{[{a_2}^{\prime}(\tau)+{a_3}^{\prime}(\tau)
c_1(\tau)]}d\tau}}}\over \sqrt{m_{0}\omega_0 {a_3}^{\prime}(t)}}\times$ \\
  & $\,\,\,$   &                                & $\quad (m_0 \omega_0
\int_0^t{{a_3}^{\prime}(\tau) e^{\int_0^{\tau}{[{a_2}^{\prime}
(\mathaccent126{\tau})+{a_3}^{\prime}(\mathaccent126{\tau})
c_1(\mathaccent126{\tau})]}d\mathaccent126{\tau}}}d\tau + i )  $ \\
  & $\,\,\,$   &                                &             \\ \hline
  &            &                                &             \\
3 & \cite{kn:ji,kn:kamke}
               & $\ddot{f}+{\dot{m(t)}\over m(t)}\dot{f}+\omega^2(t) f = 0

                                          $ & $
                             f(t)= {\epsilon(t) / \sqrt{m(t)}} $ \\
  & $\,\,\,$   &                                &             \\ \hline
  & $\,\,\,$   &                                &             \\
4 & \cite{kn:cofluctuant,kn:piza}
               & $ \dot{\sigma}={\partial <H_1>\over \partial\Pi}=
2 b \sigma + 2 a \Pi ,                        $ & $ \sigma=\sqrt{\hbar a(t)}
\epsilon(t) e^{-i\int_{0}^{t}{d\tau\over \epsilon^*(\tau)\epsilon(\tau)}},
                                                               $ \\
  &            & $ \dot{\Pi}=-{\partial <H_1>\over\partial\sigma}=
                                              $ & $ \Pi=\sqrt{\hbar\over a(t)}
\left[ {1\over2}(\dot{\epsilon}(t)-{i\over \epsilon^*(t)})-
({1\over4}{\dot{a(t)}\over a(t)}- b(t))\right]\times  $ \\
  &            & \qquad $-2 c\sigma - 2 b\Pi +
{\hbar^2 a\over 2}{1\over {\sigma}^3}         $ & $\qquad e^{-i\int_{0}^{t}
                  {d\tau\over \epsilon^*(\tau)\epsilon(\tau)}} $ \\
  & $\,\,\,$   &                                &             \\ \hline
\end{tabular}
\end{center}\ \


\begin{thebibliography}{99}
\bibitem{kn:lewis} H.R. Lewis, W.B. Risenfeld, J.Math. Phys., v.10,
 p.1458 (1969);
\bibitem{kn:trifonov1} D.A.Trifonov, Coherent States and Uncertainty Relation,
                       Phys. Lett. 48A, no.3, pp.165-66 (1974);
\bibitem{kn:trifonov2} D.A.Trifonov, Coherent States of Quantum Systems,
                       Bulgarian J. Phys, v.2, no.4, pp.303-311 (1975),\\
D.A.Trifonov, Coherent States and Evolution of
                       Uncertainty Products, Preprint ICTP IC/75/2 (1975).\\
The quantum fluctuations $\sigma_q^2$ and $\sigma_p^2$ should be
replaced in their expressions in formulas (18) in the first reference
(respectively in (15) in the second reference).
Note that the difference between these formulas and the first
two expressions in (\ref{eq:cov1}) in our article is due to the different
definitions of the Hamiltonian's coefficients;
\bibitem{kn:trifonov3} I.A.Malkin, V.I.Manko, D.A.Trifonov,
                    J.Math. Phys., v.14, no.5, p.576 (1973);\\
I.A.Malkin, V.I.Manko, D.A.Trifonov,
                   Phys.Rev. D, v.2, no.8, pp.1371-85 (1970);\\
I.A.Malkin, V.I.Manko, D.A.Trifonov,
                      Nuovo Cimento A, v.4, p.773 (1971);
\bibitem{kn:holz} A. Holz, Nuovo Cimento Lett. A, v.4, p.1319 (1970)
\bibitem{kn:lo1} C.F.Lo, Phys.Rev. A, v.43, no.1, pp.404-409 (1991);
\bibitem{kn:lee} H.W.Lee, Phys.Lett. A, v.153, no.4,5, pp.219-223 (1991);
\bibitem{kn:cheng} C.M.Cheng, P.C.Fung,
J. Phys. A, v.21, no.22, p.4115-4131 (1988);
\bibitem{kn:lo2} C.F.Lo, Nuovo Cimento B, v.105, no.5, pp.497-506 (1990);\\
C.F.Lo, J. Phys. A, v.23, p.1155 (1990);\\
C.F.Lo, Nuovo Cimento D, v.13, p.1279 (1991);\\
C.F.Lo, Europhys. Lett., v.24, p.319 (1993);\\
C.F.Lo, Nuovo Cimento, ser2, no.9, p.1015 (1995);
\bibitem{kn:agarwal} G.S.Agarwal, S.A.Kumar,
Phys.Rev. Lett., v.67, no.26, pp.3665-68 (1991);
\bibitem{kn:brown} L.S.Brown,
Phys.Rev. Lett., v.66, no.5, pp.527-529 (1991);
\bibitem{kn:paul} W.Paul, Rev. Mod. Phys., v.62, no.3, pp.531-540 (1990);
\bibitem{kn:li1} F.L. Li, Phys.Lett. A, v.168, p.400 (1992);
\bibitem{kn:li2} S.J.Wang, F.L.Li, A.Weiguni,
Phys.Lett. A, v.180, pp.189-96 (1993);
\bibitem{kn:li3} F.L. Li, S.J.Wang, A.Weiguny, D.L.Lin,
 J. Phys. A, v.27, pp.985-92 (1994);
\bibitem{kn:brito} A.L. de Brito, A.N.Chaba, B.Baseia,
Phys.Rev., A, v.52, no.2, pp.1518-24 (1995);
\bibitem{kn:ji} J.-Y. Ji, J.K.Kim, S.P.Kim,
Phys.Rev., A, v.51, no.5, pp.4268-71 (1995);\\
H.-C. Kim, M.-H.Lee, J.-Y. Ji, J.K.Kim,
Phys. Rev. A, v.53, no.6, pp.3767-72 (1996);
\bibitem{kn:seleznyova} A.N. Seleznyova,
Phys. Rev. A, v.51, no.2, pp.950-959 (1995);
\bibitem{kn:piza} A.F.R. de Toledo Piza,
Phys. Rev. A, v.51, no.2, pp.1612-1616 (1995);
The second pair variables in this work relevant to our, are:
\{ $Q= \sigma, P= \Pi$ \}.
\bibitem{kn:lai} Y.-Z.Lai, J.-Q.Liang, H.J.W. M{\" u}ler-Kirtsten,
J.-G.Zhou, Phys. Rev. A, v.53, no.5, pp.3691-93 (1996);
\bibitem{kn:gunther} N.J.Gunther, P.G.Leach,
J. Math. Phys., v.18, no.4, p.572 (1977);\\
A.K.Rajagopal, J.Marshal, Phys. Rev.A, v.26, p.2977 (1982);\\
J.Hartley, H.Ray, Phys. Rev.D, v.25, p.382 (1982);\\
X.Jing-Bo, Q.T.Zheng, G.X.Chun,
Europhys. Lett., v.15, no.2, p.119 (1991);\\
J.B.Xu, T.Z.Quang, X.C.Gao, Phys. Lett.A, v.159, p.113 (1991);
\bibitem{kn:angelow} A.Angelow, D.A.Trifonov, Schr{\" o}dinger
Covariance States in Anisotropic Waveguide, Preprint ICTP, IC/95/44
Trieste, Italy, (1995);
\bibitem{kn:cofluctuant} Cofluctuant ( or Covariance ) states with a
non-zero cofluctuation are a sub-class of Schr\"odinger minimum uncertainty
states, generalized squeezed states, two-photon coherent states e.c.
Equivalence between them has been proved for first time in:\\
D.A.Trifonov, On the stable evolution of squeezed and
correlated states, J. Sov. Laser Research, v.12, no.5, pp.414-420 (1991)
\bibitem{kn:schrodinger} E. Schrodinger, Um Heisen\-bergschen
un\-schar\-fe\-prinzip, Son\-der\-aus\-gabe aus den sit\-zungs\-berich\-ten
der preus\-sischen  akademie der wissen\-schaften,
Phys.-Math. Klasse, pp.348-356 (1930), v.XIX ;
\bibitem{kn:walls1} M.D.Levenson, R.M.Shelby, A.Aspect, M.Reid, D.F.Walls,
 Phys. Rev. A, v.32, no.3, pp.1550-62 (1985);
\bibitem{kn:drummond} P.D.Drummond, S.J.Carter,
J.O.S.A., vol.4, no.10, pp.1565-73 (1987);\\
P.D.Drummond, Phys.Rev., A, v.42, no.11, pp.6845-57 (1990);
\bibitem{kn:yariv} A.Yariv, Quantum Electronics, John Wiley and Sons,
Inc., 3rd Ed., New York (1988);
J.Perina,  Quantum  statistic  of  linear  and  nonlinear  optical
phenomena, D. Reidel Publishing Company, Boston (1984);
\bibitem{kn:nikolov} I.Zlatev, A.Nikolov, Theoretical Mechanics, v.1,
                     N.I., Sofia (1981);
\bibitem{kn:kamke} E.Kamke, Differentialgleichungen, L\"{o}sungsmethoden
und L\"{o}sungen, Gew\"{o}hnliche differentialgleichungen, vol.I., Leipzig,
Akademische Verlagsgesellschaft (1959)

\end{thebibliography}
\end{document}